\documentclass{emulateapj}
\usepackage{color,psfrag}

\begin{document}

\def\deg{^{\circ}}

\newcommand{\psrae}{\mbox{J0737$-$3039A}}
\newcommand{\psra}{\mbox{J0737$-$3039A }}
\newcommand{\psrbe}{\mbox{J0737$-$3039B}}
\newcommand{\psrb}{\mbox{J0737$-$3039B }}
\newcommand{\be}{\begin{eqnarray}}
\newcommand{\ee}{\end{eqnarray}}
\newcommand{\ra}{{\bf r_A}}
\newcommand{\rb}{{\bf r_B}}
\newcommand{\xbz}{{\rm x_B(0)}}
\newcommand{\ybz}{{\rm y_B(0)}}
\newcommand{\va}{{\bf V_A}}
\newcommand{\vb}{{\bf V_B}}
\newcommand{\vao}{V_{AO}}
\newcommand{\vbo}{V_{BO}}
\newcommand{\sx}{s_x}
\newcommand{\sy}{s_y}
\newcommand{\ta}{t_A}
\newcommand{\tb}{t_B}
\newcommand{\tap}{t_{AP}}
\newcommand{\tbp}{t_{BP}}
\newcommand{\kms}{km~s$^{-1}$ }
\newcommand{\kmse}{km~s$^{-1}$}
\newcommand{\viss}{{\bf V_{\rm ISS}}}
\newcommand{\siss}{S_{\rm ISS}}
\newcommand{\tiss}{T_{\rm ISS}}
\newcommand{\vpar}{V_{\rm par}}
\newcommand{\vperp}{V_{\rm perp}}
\newcommand{\ar}{A_r}

\title{Probing the Eclipse of J0737$-$3039A with Scintillation}
\author{W. A. Coles\altaffilmark{1}, M.\ A.\ 
McLaughlin\altaffilmark{2}, B. J. Rickett\altaffilmark{1}, A. G. 
Lyne\altaffilmark{2}, \&  N. D. R. Bhat\altaffilmark{3}}
\altaffiltext{1}{University of California, San Diego}
\altaffiltext{2}{Jodrell Bank Observatory, University of Manchester, 
Macclesfield, Cheshire, SK11 9DL, UK}
\altaffiltext{3}{Massachusetts Institute of Technology, Haystack 
Observatory, Westford, MA 01886}

\begin{abstract}

We have examined the interstellar scintillations of the pulsars in the
double pulsar binary system. Near the time of the eclipse of pulsar A
by the magnetosphere of B, the scintillations from both pulsars should
be highly correlated because the radiation is passing through the same
interstellar plasma. We report confirmation of this effect using 820
and 1400 MHz observations made with the Green Bank Telescope. The
correlation allows us to constrain the projected relative position of
the two pulsars at closest approach to be $4000 \pm 2000$~km,
corresponding to an inclination that is only $0.29\pm 0.14^\circ$
away from 90$^\circ$.
It also produces a two-dimensional map of the spatial correlation of the
interstellar scintillation. This shows that the interstellar medium in
the direction of the pulsars is significantly anisotropic. When this
anisotropy is included in the orbital fitting, the transverse velocity
of the center of mass is reduced from the previously published value
of 141~$\pm 8.5$~\kms to 66~$\pm15$~\kmse.

\end{abstract}

\keywords{pulsars: general -- pulsars: individual (J0737$-$3039) -- ISM:
general -- binaries: general}

\clearpage

\section{Introduction} \label{sec:intro}

The two pulsars in the recently discovered double pulsar binary
system, ``A'' and ``B'', have periods of 23~ms and 2.8~s,
respectively. They are in a highly relativistic orbit with period of
only 2.4~hrs and eccentricity of 0.088 and have an average separation of only
2.9~lt-s \cite{burgay03,lyne04}. The
phenomenology exhibited by this system is extremely rich.  Pulsar A is
eclipsed when it passes behind pulsar B
\cite{lyne04,kaspi04}. Modulation of the eclipse with the period of B
 is consistent with the eclipse being caused by synchrotron absorption by 
the plasma surrounding the magnetosphere of B \cite{mcl04b}.
Throughout the orbit, the flux density of the B
pulsar varies dramatically and systematically, indicating significant
interaction between the two pulsars \cite{ramach04,mcl04a}. Only $\sim$ 0.5~kpc away, and
with orbital velocities reaching over 300 \kmse, this system is an
ideal laboratory for studies of general relativity, binary kinematics
and, in the case of this paper, the binary orbital geometry and the
interstellar medium.

Pulse timing measurements provide many of the parameters of the
system, including the exact masses of the two pulsars, with exquisite
precision.  From measurements of the Shapiro delay, the inclination of
the system can be constrained to be 87$\deg\pm3\deg$
\cite{lyne04}. Measurements of the time scale of the interstellar
scintillation (ISS) over an orbit can be used to estimate the orbital
inclination, the longitude of periastron, and the transverse velocity
of the center of mass of the system. This technique was first
suggested by Lyne \& Smith (1982) and has been applied to PSR~B0655+64
\cite{lyne84} and PSR~J1141$-$6545 (Ord, Bailes \& van Straten
2002). In general, there are two solutions for the orbital inclination
that fit the data equally well, but one may fit the expected mass of
the pulsar better.  In the case of the double pulsar system, the
masses of A and B are already accurately determined through timing,
removing this ambiguity.

Ransom et al. (2004) (hereafter R04) applied the technique to the double pulsar
system, calculating an inclination of $88.7\deg\pm0.9\deg$ and
transverse velocities of the center of mass parallel and perpendicular to 
the orbital line of nodes in the plane of the sky of $\vpar=96.0 \pm 3.7$~\kms and
$\vperp=103.1\pm7.7$~\kmse, respectively. This calculation relied on the 
implicit assumption that the turbulence in the interstellar medium (ISM) is 
isotropic.  If the turbulence is anisotropic the transverse velocity estimated by
this method depends strongly on the axial ratio and orientation of the
anisotropy and can be much lower than for isotropic turbulence.
Note that neither this method nor timing can resolve the {\it sense} of the 
inclination, i.e. whether the angular momentum vector points towards or away from 
the Earth. Furthermore neither method can resolve the direction of the parallel 
velocity with respect to the angular momentum vector.

In this paper, we examine the correlations of the A and B pulsars'
scintillations near the time of the A eclipse, i.e. when the lines of
sight are closest to one another. We show how these
correlations constrain the orbital geometry more tightly than was
previously possible and partially resolve the ambiguity in the sense of the
inclination and the direction of the perpendicular velocity. 
We also show that the ISM is significantly anisotropic and
that the transverse velocity of the center of mass is much smaller than
was estimated with an isotropic model of the ISM. 

\section{Observations and Analysis} \label{sec:obsandanal}

Observations of the double pulsar system with the 100-m Green Bank
Telescope took place in December 2003 and January 2004 under an
``Exploratory Time'' proposal as reported by R04. The
820-MHz data used for this analysis were acquired with the GBT
Spectrometer SPIGOT card with a sampling time of 40.96~$\mu$s on each of
1024 frequency channels covering a 50-MHz bandwidth. The 1400-MHz data
were acquired using the Berkeley-Caltech Pulsar Machine (BCPM) using a
sampling interval of 72~$\mu$s on each of 96 channels covering a 96-MHz
bandwidth. A total of 5 hours of data at 820 MHz and 6.1 hours of data 
at
1400 MHz were obtained.

Our first step in the analysis of these data was to create dynamic spectra
for each pulsar in each frequency band. In order to do this, we folded each
frequency channel modulo the pulsar period using the ephemeris from
Lyne et al. (2004) and freely available software tools (Lorimer et
al. 2000). Individual frequency channels were shifted with respect to
each other using a dispersion measure (DM) of 48.9~cm$^{-3}$~pc (Lyne
et al. 2004). Individual profiles were added so that the time resolution
was 10~s at 1400~MHz and 5~s at 820~MHz. 
We formed on-pulse spectra by integrating over an ``on-pulse'' window and
subtracted off-pulse spectra formed by integrating over an equivalent ``off-pulse''
window.
We did not attempt to correct for variations in the system gain
over the bandpass because such corrections increased the noise and the gain
variations did not appear to bias the results.
The dynamic spectra are essentially the same as those shown by R04.

In order to estimate the time scale we first split the dynamic
spectrum of A into segments of length 5.3~min and 10~min at 820 and
1400 MHz respectively, subtracted the mean from each spectrum, 
and then computed the temporal autocorrelation
function $\rho(t)$ of each segment. 
We then fit a theoretical model
beginning at the second point to determine the time scale $\Delta
t_d$, as defined in Cordes (1986). We used the theoretical form for a
Kolmogorov scattering medium $\rho(t) = \exp[-(t/\Delta t_d)^{5/3}]$
rather than the commonly used Gaussian function $\rho(t) = \exp[-(t/\Delta t_d)^{2}]$. 
The rms error in the $\Delta t_d$ estimates is much smaller
at 820 MHz because the 820 MHz spectra have 10 times as many degrees
of freedom as the 1400 MHz spectra.

Normally, scintillation is thought of as primarily a temporal process
$I(t)$, i.e. ``twinkling'' caused by the motion of a spatial diffraction
pattern $I({\bf r})$ past the observer with transverse velocity $\viss$.
If the diffraction pattern is isotropic with spatial scale $\siss$, the
time series will have a time scale $\tiss = \siss/|\viss|$. The spatial
scale can be estimated from the bandwidth and the distribution of
turbulence along the line of sight \cite{cr98}, so that one may estimate
the velocity as $|\viss| = \siss/\tiss$. The transverse velocity of the
center of mass ($\vpar,\vperp$) and $\siss$ are derived by fitting a model to the
measured $|\viss(\phi)|$, where $\phi$ is the orbital phase. As noted by
R04 the result is calibrated relative to the precisely
known orbital velocity of the A pulsar. However, if the ISM is
anisotropic, $\siss$ depends on the direction of $\viss$ and the model
must be modified accordingly (see \S\ref{sec:aniso}). Our estimate of
($\vpar,\vperp$) using an isotropic model matches that of R04
within their quoted errors.

\section{Cross-correlation of Pulsars A and B}

Near the time of the eclipse of pulsar A, the lines of sight to the
two pulsars will pass through nearly the same ISM and their
scintillations should therefore be correlated. The pulsars sample the
ISM at their projected locations $\ra(\ta) = \ra(0) + \va \ta$ and
$\rb(\tb) = \rb(0) + \vb \tb$ which, of course, depend on the times of
observations.  $\va$ and $\vb$ are the vector sums of the transverse
center of mass velocity ($\vpar,\vperp$) with the transverse orbital
velocities $\vao$ and $\vbo$.  We can regard $\va$ and $\vb$ as essentially
constant near the time of the eclipse. We define $\ta$ and $\tb$ with
respect to the time of the superior conjunction of A (where A is at
least partially eclipsed). Likewise, $\ra$ and $\rb$ are
defined with respect to the position of A at $\ta = 0$.  We define
our coordinates as parallel and perpendicular to the line of nodes.
Since the inclination is so close to 90 degrees, 
this is essentially parallel to pulsar B's orbital velocity.
The location of B at superior conjunction is $\ybz$.
The
spatial cross correlation $\rho({\bf s} = \ra-\rb)$ cannot be computed
as a time average because the baseline ${\bf s}$ varies rapidly with
time. However the averaging can be done over frequency because ISS
also modulates the spectrum, and the spectral modulation 
is independent of velocity.

\begin{figure}[h]
\psfrag{Ta(s)}[][]{$\it t_A$ (s)}
\psfrag{Tb(s)}[][]{$\it t_B$ (s)}
\includegraphics[scale=0.50]{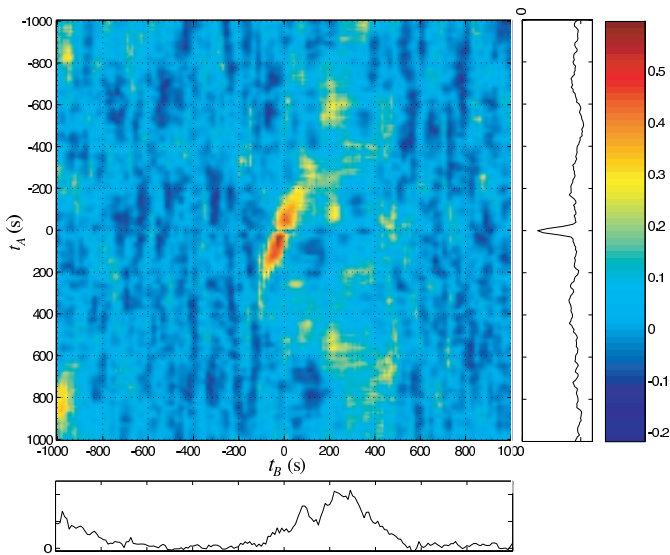}
\caption{The cross correlation $\rho(\ta,\tb)$ averaged over the three
eclipses measured.  The flux densities of A and B are shown as marginal
distributions, showing clearly the eclipse of A and the strong orbital
variation of B. The correlation has not been corrected for system
noise.}
\label{fig:tcor1400}
\end{figure}

In Figure~\ref{fig:tcor1400}, we present the cross-correlation
$\rho(\ta,\tb)$ of the dynamic spectra of A and B at 1400~MHz. This is
an average of all three eclipses in the data set. The spectra were
smoothed over three samples (30 s) in time, retaining 10 s sampling,
in order to optimize the signal to noise ratio. The correlation is
normalized by the square root of the product of the total variance
from A and B, i.e. including noise.  The marginal distributions show
the fluxes of A and B as a function of time.  The intensity of B is
very low more than 200 s before the eclipse, making it impossible to
correct the normalization for system noise over the range of
($\ta,\tb$) plotted in Figure~\ref{fig:tcor1400}. We corrected the
correlation for noise in the region where the flux of B is significant
($\tb > -200$ s)
and fitted an elliptical Kolmogorov model to the data. The peak correlation
was 81\% at $\tap=33.4\pm 2.8$~sec (i.e.\ after the eclipse)
and $\tbp=-22.7\pm 1.2$~sec (i.e.\ before the eclipse). The noise correction 
is uncertain because the noise is not stationary and we believe that the peak
correlation is consistent with 100\%. At this peak the A and B trajectories 
$\ra(\ta)$ and $\rb(\tb)$ intersect as shown in Figure~2.

The intersection equation $\ra(\tap) = \rb(\tbp)$ yields 
$\va \tap = \vb \tbp + \ybz$.
The x component provides a solution for the center of mass velocity
component $\vpar = (\vao \tap + \vbo \tbp)/(\tap - \tbp)$.
Using the known orbital speeds at the time of the eclipse
($\vao=329$, $\vbo=352$ \kms), we directly obtain $\vpar=53\pm 17$ \kms. This
is clearly inconsistent with $\vpar = 96\pm4$ \kms obtained by R04.
We will show later that this inconsistency is due to anisotropic
scattering in the interstellar plasma. The y component of the intersection
equation yields $\vperp = \ybz/(\tap-\tbp)$.
As $(t_{\rm AP}-t_{\rm BP})$ is positive, we know that $\vperp$ is in the same 
direction as $\ybz$, but we cannot determine if they are parallel or anti-parallel
to the angular momentum. Using the value $\vperp = 103\pm7.7$ \kms determined by
R04 we find $\ybz = 5780$ km.
The trajectories $\ra(\ta)$ and $\rb(\tb)$ shown in Figure~2 are calculated using
the velocities determined by R04 assuming isotropic scattering and this value of $\ybz$.
The alternative track for $\rb(\tb)$ shown without a dotted line 
corresponds to the ambiguity in the sense of the orbital inclination. 
The inclination is 0.41$^\circ$ from 90$^\circ$ if $\ybz$ is correct.
However anisotropic scattering also modifies $\vperp$ as we will show
later, and this reduces $\ybz$. While the pulsars will again sample similar paths through
the ISM when B passes behind A, the emission from B is very weak at
this crossing and we cannot repeat this analysis for that
part of the orbit.

\begin{figure}[h]
\includegraphics[scale=0.45]{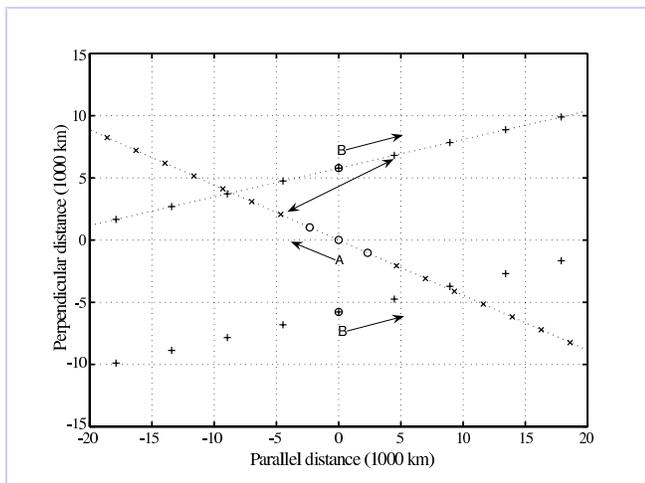}
\caption{The trajectories of the A and B pulsars referenced to the
location of the A pulsar at the time of its eclipse.
The sample spacing is 10~s. The duration of the eclipse is indicated by
open circles on the A trajectory, otherwise denoted by crosses connected
by a dotted line.
The plus signs connected by a dotted line represent the B
trajectory for an inclination of
90.41$\deg$, corresponding to an $\ybz$ of 5780~km.
The plus signs with no dotted line show the B trajectory for the opposite
sense of inclination. To illustrate the mapping more clearly, the
baseline ${\bf s} = {\bf r}_A(20) - {\bf r}_B(10)$
where $(\ta,\tb) = (20,10)$ is indicated by the double-ended arrow. }
\label{fig:traj}
\end{figure}

\begin{figure}[ht]
\psfrag{Xa - Xb \(km\)}[c][]{${\rm x_A - x_B}$ ($10^4$ km)}
\psfrag{Ya - Yb \(km\)}[c][]{${\rm y_A - y_B}$ ($10^4$ km)}
\includegraphics[scale=0.50]{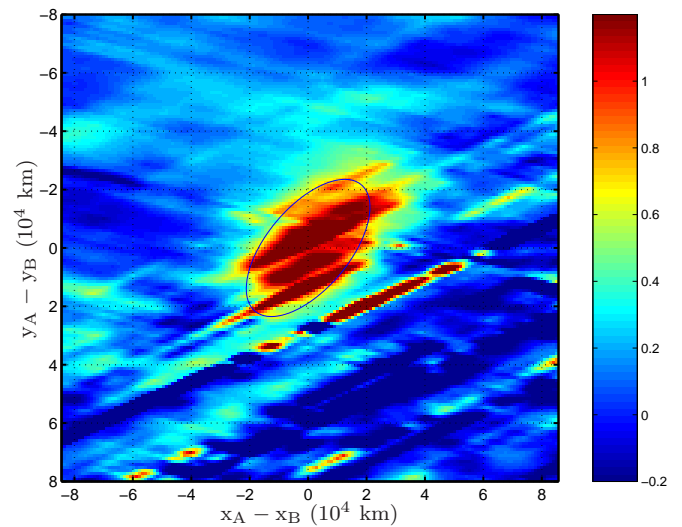}
\caption{The apparent spatial correlation $\rho(s_x,s_y)$ at 1400~MHz, where
$({\rm x_A, y_A})$ and $({\rm x_B, y_B})$ are the parallel and
perpendicular
components of ${\bf r}_A$ and ${\bf r}_B$, respectively. The
mapping velocities are those determined from an isotropic scattering
model. An ellipse with an axial ratio of 2 with the major axis at
--50$\deg$ is drawn over the correlation to guide the eye. Here $\ybz
= 4000$~km is used because it provides a more symmetric result
than the 5780~km shown in Figure~\ref{fig:traj}.}
\label{fig:scor1400}
\end{figure}

The temporal correlation in Figure~\ref{fig:tcor1400} can be mapped
into a spatial correlation $\rho({\bf s} = \ra(\ta) - \rb(\tb))$
if the velocities are known. This will allow us to examine the
isotropy assumption directly. The spatial correlation resulting from the
trajectories shown in Figure~\ref{fig:traj} is displayed in 
Figure~\ref{fig:scor1400}. Here we have displayed a more limited
range than in Figure~\ref{fig:tcor1400} and we have corrected the
normalization of the correlations for system noise over the entire plot. 
This reduces the bias due to system noise but it creates a diagonal 
stripe of spurious correlation running from (0, $-30000$ km) to 
(60000 km, 0) where the flux of pulsar B is very low 
(so the correction factor is very large).
When $\ybz$ is chosen correctly, the (noise-corrected) spatial
correlation should be symmetrical around the origin. The value of
$\ybz$ which best centers the correlation is 4000~km. Note
that because the spatial scale of the ISS is larger than that of
the eclipse, the spatial correlation is not badly distorted by A's
eclipse.  This apparent correlation is clearly 
anisotropic with an axial ratio near 2.  
However, since the mapping used the center of
mass velocity obtained from an isotropic scattering model, this result
is not self-consistent.  Consequently, in \S\ref{sec:aniso} we repeat
the orbital time scale fitting including anisotropy. 

\begin{figure}[ht]
\psfrag{Tb \(s\)}[][]{$\it t_B$ (s)}
\psfrag{Ta \(s\)}[][]{$\it t_A$ (s)}
\includegraphics[scale=0.45]{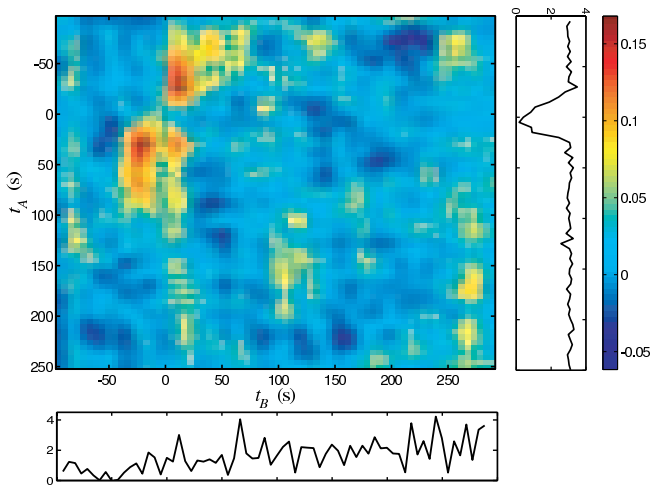}
\caption{The average correlation $\rho(t_A,t_B)$ for the three eclipses
at
820~MHz.  The sample interval is 5~s. The origin is offset because the
first eclipse was very close to the start of the observation. The
spatial
scale of the ISS is much smaller than at 1400~MHz and is comparable to
the
scale of the eclipse, so that the correlation is heavily distored by the
eclipse, and perhaps by propagation through the B magnetosphere.
However,
the peak remains in the lower left, as at 1400~MHz (Figure 1).}
\label{fig:tcor820} 
\end{figure}

\begin{figure}[ht]
\psfrag{Xa - Xb \(km\)}[c][]{${\rm x_A - x_B}$ ($10^4$ km)}
\psfrag{Ya - Yb \(km\)}[c][]{${\rm y_A - y_B}$ ($10^4$ km)}
\includegraphics[scale=0.50]{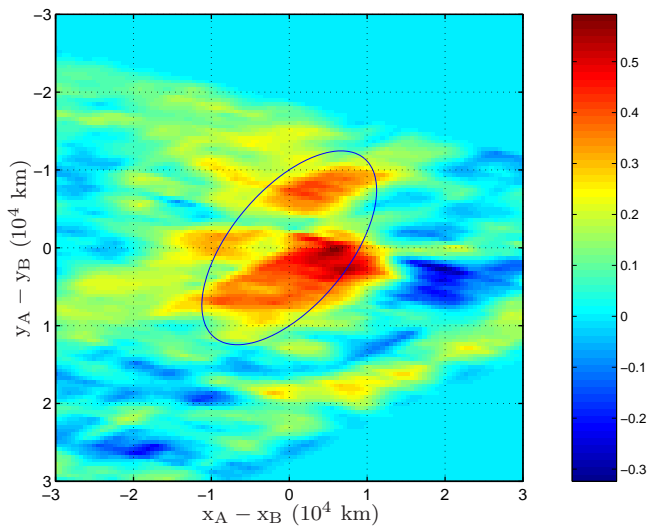}
\caption{The spatial correlation $\rho(s_x,s_y)$ at 820~MHz. The edge
of the transformation window can be seen because the first eclipse was
very close to the start of the observations.  This correlation has
been corrected for 95\% of the system noise. Full correction produced
high errors during the eclipse. 
The ellipse drawn over the 1400~MHz data has been scaled down
by a factor of 53\% and over-plotted here.}
\label{fig:scor820}
\end{figure}

In Figure~\ref{fig:tcor820}, we present the temporal correlation at
820 MHz, averaged over three eclipses. As at 1400 MHz, the spectra
were smoothed over three sample intervals but the original 5 s
sampling was retained.  The spatial scale of ISS scales as 
$\lambda^{-2/\alpha}$ (eqn 44 of Rickett, 1977), where $\alpha = 5/3$
for a Kolmogorov scattering medium. Thus the scale
at 820 MHz is smaller than at 1400~MHz by a factor of 
$(820/1400)^{1.2} = 0.53$, making it more
comparable with the size of the eclipse.  The raw correlations are
lower than at 1400 MHz, but the estimation error is also lower. When
the normalization is corrected for system noise the peak correlation
reaches 100\%. The spatial correlation at 820~MHz is shown in
Figure~\ref{fig:scor820}, using the same ($\vpar,\vperp$) as for
Figure ~\ref{fig:scor1400}.  The correlation is also anisotropic,
but the axial ratio is somewhat smaller than at 1400 MHz. We attribute
the differences to the greater distortion by the eclipse at 820 MHz.
To make this correlation symmetric about the origin we reduced $\ybz$
to 3000~km. It is possible that the difference between this value of
$\ybz$ and that of 4000~km at 1400 MHz is due to differential
refraction in the magnetosphere of B, but the difference is comparable
with our estimate of the error.  Allowing for this possibility we
estimate that $\ybz = 4000 \pm 2000$~km which corresponds to an
inclination of $0.29^\circ\pm0.14^\circ$ away from 90$^\circ$.

\section{Anisotropy Analysis for J0737$-$3039}
\label{sec:aniso}

While the previous analysis showed that the ISS is anisotropic, the
spatial mapping was carried out with velocities determined using an
isotropic scattering model, hence the result is not
self-consistent. In this section, we therefore modify the orbital
fitting to include anisotropy and find a self-consistent spatial
mapping.

If the spatial correlation has elliptical symmetry, it can be written as
$\rho({\bf s}) = f(Q({\bf s}))$, where
$Q({\bf{s}}) = a s_x^2 + b s_y^2 +c s_x s_y$
is the quadratic form which describes the ellipse.
The form of $f$ is less important; for example, for a Kolmogorov power law,
$\rho({\bf s}) = {\rm exp}[-(Q({\bf s})/\siss^2)^{5/6}]$. 
Here, $\siss$ is the 1/$e$ spatial scale. The temporal correlation
$\rho(t)$ is determined by substitution of ${\bf s} = {\bf V} t$, giving
$\rho(t) = f(Q({\bf V}) t^2)$, where ${\bf V}$ is the
transverse velocity of the pulsar. The width $\tiss$ of the temporal
correlation is given by $Q({\bf V}) \tiss^2 = \siss^2$. In
previous work, $\siss$ has been calculated from a theoretical scattering
model, with the theoretical velocity ${\bf \va}(\phi)$ fit to the measured
$|\viss(\phi)| = \siss /\tiss(\phi)$ with an arbitrary scale factor 
(e.g. Ord et al. 2002). However, the preliminary calculation of
$\siss$ is not necessary, so we have simply included $\siss$ in the fit
instead of using a scale factor. The model we fit to the measured values
of $(1/\tiss)$ for pulsar A is $[Q({\bf \va})]^{0.5}/\siss$,
and the fitting parameters are $\vpar$, $\vperp$, and
$\siss$. The input parameters $\ar$ and $\theta$, where $\ar$ is the axial
ratio of the fitted ellipse and $\theta$ is the position angle of the
ellipse, must be given. The coefficients of the quadratic form are:
$a = {\rm cos}^2 \theta/\ar + \ar {\rm sin}^2 \theta$; $b = {\rm sin}^2
\theta/\ar + \ar {\rm cos}^2 \theta$; and $c = 2{\rm sin}\theta {\rm
cos}\theta(1/\ar - \ar)$. With this definition, $\siss$ is the geometric
mean of the major and minor axis scales.

To find a self-consistent solution we use the 1400 MHz correlations, 
which are least distorted by the eclipse, with the 820 MHz 
measurements of $\tiss(\phi)$ which have the higher signal to noise 
ratio. We then assumed an axial ratio and orientation; 
computed the resulting center of mass velocity
by fitting to the time scale vs orbital phase; used that velocity
to create a spatial correlation; and estimated the axial ratio and
orientation of that spatial correlation. When the correct axial ratio
has been assumed it will be the same as that of the spatial correlation.
We did this for a grid of axial ratio and orientation up to axial ratios 
of 10. To define a goodness of fit we used the technique
suggested by Grall et al. (1997). We plotted the assumed anisotropy
and that of the resulting spatial correlation on a modified 
Poincar\'e sphere and used the distance between them as a measure of 
the error. The best solution, near $\ar = 4$ and $\theta = -15^\circ$ 
with $\ybz = 4000$~km, is shown in Figure~6.
We find acceptable solutions for $\ar \ge 3$ with $-20^\circ \leq \theta \leq
-10^\circ$. We searched $\ar$ as large as 10, but we find it hard to
believe that such axial ratios are feasible without corroborating
evidence. The center of mass velocity is minimum at $\ar
= 4$. If we restrict the axial ratio to $3 \leq \ar \leq 6$ then $51
\leq |(\vpar,\vperp)| \leq 81$ km s$^{-1}$. 
The estimated $\vpar = 53\pm17$ \kms obtained earlier is consistent with
this result and it may be possible with more data to use $\vpar$ to 
improve the final model, but the data and analysis do not justify that
at present. Thus we conclude
that the center of mass velocity is much lower than previously thought.  

The best fit to the $\tiss(\phi)$ measurements with this anisotropy 
has  $\siss = 9900\pm300$~km at 820~MHz. Note that
this is the geometric mean of the scales on the major and minor axes.
This $\siss$ corresponds to 18700~km at 1400~MHz which agrees well with
the spatial correlation shown in Figure~6. It is not possible to compare 
$\siss$ with a theoretical calculation at present as the appropriate 
calculations for anisotropic scattering have not yet been done.

\begin{figure}[ht]
\psfrag{Xa - Xb \(km\)}[c][]{${\rm x_A - x_B}$ ($10^4$ km)}
\psfrag{Ya - Yb \(km\)}[c][]{${\rm y_A - y_B}$ ($10^4$ km)}
\includegraphics[scale=0.50]{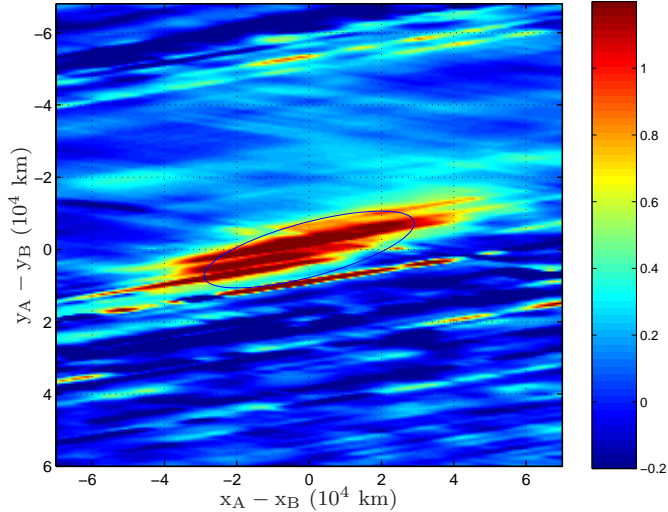}
\caption{The final
self-consistent spatial correlation $\rho(s_x,s_y)$ of the ISM in the
direction of the J0737$-$3039 system for the observations at 1400~MHz. The best-fit
ellipse, with $\ar = 4$ and $\theta = -15\deg$, is shown.}
\label{fig:finalcor}
\end{figure}

Previous detections of anisotropic interstellar scattering have
come largely from angular broadening measurements of highly
scattered objects. Measured axial ratios are
in the range 1.3 to 3.0 (Wilkinson et al. 1994;
Molnar et al. 1995; Desai \& Fey 2001;
Frail et al. 1994; Yusef-Zadeh et al. 1994;
Desai et al. 1994).  An axial ratio of at least 4:1 was inferred
from scintillation analysis of the quasar B0405--385
by Rickett at al. (2002).  Though the latter result was
identified as due to a discrete region only
15 -- 30 pc from the Earth, it confirms that localized
regions of the interstellar medium can cause
axial ratios as high as we find here. 

It may be possible to confirm the anisotropy in this system with
a more detailed analysis of the ISS using the ``parabolic arc'' 
phenomenon (Stinebring et al.\ 2001) and by extending the observations
over a year so we can observe the annual modulation caused by the
Earth's velocity. Parabolic arcs have been observed in this system and 
they provide a different view of the ISS. It is not fully independent
of the time scale analysis but it has a different dependence on
anisotropy which is the critical feature of our analysis.

\section{Anisotropy Analysis in General}

\begin{figure}[ht]
\includegraphics[scale=0.45]{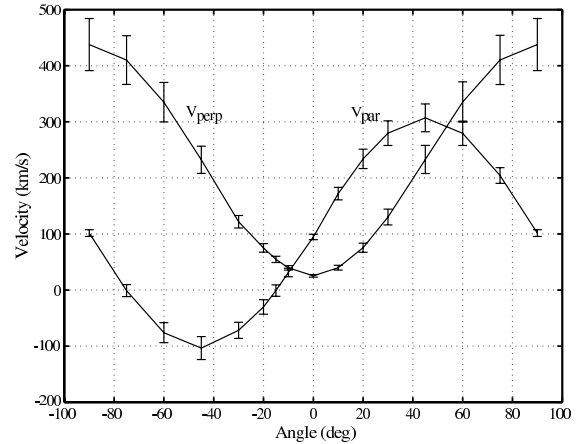}
\caption{$\vpar$ and $\vperp$ derived from fitting of an anisotropic
model to measurements
of $\viss$ for $\ar = 4$ at various angles for J0737--3039. All fits had the same qu
ality.
The 95\% confidence limits are shown.}
\label{fig:fits}
\end{figure}

In fitting the $\viss(\phi)$ measurements with anisotropic models, we
realized that, because the inclination of the orbit is so close to
90$^\circ$, $\ar$ and $\theta$ are completely degenerate parameters. The
$\viss(\phi)$ measurements themselves can be fit equally well with any
anisotropy (although with different $\siss$, $\vpar$ and $\vperp$).  The
effect of anisotropy on $\vpar$ and $\vperp$ is shown for $\ar$ = 4 in
Figure~\ref{fig:fits}.  The cause of the degeneracy can be deduced from
the equations for ${\bf \va}(\phi)$ directly
(equations 5-7 of Ord et al., 2002). The model for the 
degenerate
orbit as a function of orbital phase $\phi$ has the form ${\bf
\va}(\phi)^2 = A {\rm cos}2\phi + B {\rm sin}\phi + C$ (ignoring terms 
of the order of
the eccentricity).  Therefore, one can fit only three independent
variables. Note that, in the case of the double pulsar system, the 
spatial
correlations provide an estimate of the anisotropy and remove this
degeneracy.

In general the ${\bf \va}(\phi)^2$ model has the form ${\bf
\va}(\phi)^2 = A {\rm cos}2\phi + B {\rm sin} 2\phi + C {\rm sin}\phi +
D {\rm cos}\phi + E$ (again, ignoring terms of the order of the
eccentricity).  Thus one can fit five independent variables. In the
system J1141$-$6545, Ord et al.  (2002) fit $\siss$, $\vpar$,
$\vperp$, inclination, and the longitude of periastron. They found a
most likely center of mass speed of 115 \kmse.  We tested the effect
of anisotropy on this system by fitting the measurements of Ord et
al. (2002) and reproducing their result. Then, holding the longitude
of periastron constant, we introduced an axial ratio and fit for the
remaining parameters. We found a wide range of fits for $1.5 < \ar <
3$, all of which were indistinguishable from the best isotropic
fit. All of the fits we tried had greatly reduced $\vpar$ and
$\vperp$. For example, with $\ar = 2.1$, we obtain $\theta = 0$,
inclination = $75^{\deg}$, $\vpar = 20.5$~\kms and $\vperp =
17.1$~\kmse.

Although we do not know the value of the anisotropy for J1141$-$6545,
this is an important result, because it greatly weakens the case for
high transverse velocities in binary systems. Observationally it makes
the problem of measuring anisotropy in the ISM of more immediate
interest.

\section{Conclusions} \label{sec:conclusions}

The cross correlation between the ISS of the two pulsars in the double
pulsar system has been measured. It 
tightly constrains the geometry of the eclipse.
We estimate the projected relative distance between the two pulsars at
eclipse to be $4000 \pm 2000$~km.  McLaughlin et al. (2004b)
have shown that the eclipse duration varies, depending strongly on B
pulse phase. However, if we take the lateral extent of the eclipse to be
roughly 18,600~km (derived from the maximum 680~\kms relative velocity 
of
the two pulsars and the average eclipse duration of 27~s), the eclipse
must occur at an average radial distance from B of about 9000~km, only 
7\%
of the 130,000~km radius of the light cylinder of B. This is further
evidence that the relativistic wind from A has blown away much of the
magnetosphere of B.

Our inclination estimate, 0.29$^\circ \pm 0.14^\circ$ away from 90$^\circ$,
is consistent, within the errors, with the
inclination determined from the Shapiro delay. However, the errors on the
timing-derived inclination are decreasing significantly with continued
observations. If the estimate of inclination determined through the ISS
measurements differs from that measured from timing, we may be able to use
the measurements of the ISS correlation at different frequencies to
estimate the refraction of pulsar A by the magnetosphere of B near the
eclipse and determine the density gradient in the magnetosphere.
Note that the our inclination estimate is sufficiently small that gravitational
lensing may be important (Lai \& Rafikov 2004). This effect would bias both
the ISS and Shapiro delay estimates of inclination.

The correlations show that the ISS in the direction of the system is
anisotropic. When this anisotropy is included in the orbital analysis, the
transverse velocity of the center of mass is reduced from $141 \pm
8.5$~\kms to $66 \pm 15$~\kmse.  We also see that a modest anisotropy in
other binary systems, such as J1141$-$6545, can greatly reduce the implied
center-of-mass velocity. This decreases the need to invoke large kick
velocities and/or high precollapse core masses in double neutron star
formation scenarios, as suggested by R04. In fact, as
discussed by Piran \& Shaviv (2004), the low eccentricity of the system,
coupled with its location close to the Galactic plane, suggest a B
progenitor mass less than $2~M_\odot$, and most likely around
$1.45~M_\odot$. This implies a non-standard, possibly white dwarf,
progenitor. Willems \& Kalogera (2004), who considered the orbital
evolution of the system due to gravitational radiation and the orbital
dynamics of asymmetric supernova explosions, likewise have difficulty
explaining kick velocities of less than 60 km~s$^{-1}$ given mass ranges
derived by assuming a Helium star progenitor and current models of Helium
star evolution. Another way we may be able to confirm the low center of mass
velocity of the system is through a measurement of the proper motion.
Given the closeness of this system, such a measurement should be possible
within a year. This will allow us to place better constraints on the kick
velocity, the progenitor mass of B and evolutionary scenarios, and will
also allow us to test our main conclusion of a significantly anisotropic
ISM.

\acknowledgments

We are grateful to NRAO for making these data available through the
Rapid Response Science program and to Scott Ransom for taking the
data. We thank Carlos Nava for his help estimating the time scales and
fitting them with an anisotropic scattering model and Duncan Lorimer
for writing the publically available SIGPROC package. The National
Radio Astronomy Observatory is facility of the National Science
Foundation operated under cooperative agreement by Associated
Universities, Inc.  Coles and Rickett acknowledge partial support
from the NSF under grant AST 9988398.

{}
\end{document}